\documentclass[superscriptaddress,showpacs,twocolumn,prl,floatfix]{revtex4}
\usepackage{amsmath}
\usepackage{amssymb}
\usepackage{graphicx}
\usepackage{dcolumn}
\usepackage{bm}
\usepackage{ulem}

\begin{document}

\title{Beating of Friedel oscillations induced by spin-orbit interaction}
\author{S. M. Badalyan}
\email{Samvel.Badalyan@physik.uni-regensburg.de}
\affiliation{Department of Physics, University of Regensburg, 93040 Regensburg, Germany}
\affiliation{Department of Radiophysics, Yerevan State University, 1 A. Manoukian St.,
Yerevan, 375025 Armenia}
\author{A. Matos-Abiague}
\affiliation{Department of Physics, University of Regensburg, 93040 Regensburg, Germany}
\author{G. Vignale}
\affiliation{Department of Physics and Astronomy, University of Missouri - Columbia,
Missouri 65211, USA}
\author{J. Fabian}
\affiliation{Department of Physics, University of Regensburg, 93040 Regensburg, Germany}
\date{\today}

\begin{abstract}
By exploiting our recently derived exact formula for the Lindhard polarization function in 
the presence of Bychkov-Rashba (BR) and Dresselhaus (D) spin-orbit interaction (SOI),
we show that the interplay of different SOI mechanisms induces highly anisotropic 
modifications of the static dielectric function. We find that under certain circumstances 
the polarization function exhibits doubly-singular behavior, which leads to an intriguing 
novel phenomenon, beating of Friedel oscillations. This effect is a general feature of 
systems with BR+D SOI and should be observed in structures with a sufficiently strong SOI.
\end{abstract}

\pacs{72.25.Dc, 72.10.-d, 73.63.Hs, 73.21.Fg}
\maketitle

Spin-orbit interaction (SOI) is of great interest for spintronic
applications~\cite{zfds,fmesz}. Electron spin is not conserved in the
presence of SOI, which allows for purely electric manipulation of spins~\cite%
{grundler,nitta}. In conjunction with other carrier scattering mechanisms,
SOI leads to intriguing novel phenomena. One of the major findings is the
detection of the spin Hall effect~\cite{murakami,kato,wunderlich,sih},
predicted long ago as an outcome of the interplay between SOI and
electron-impurity scattering~\cite{dyakonov}. In turn, electron-electron
scattering mediates mutual transformations of spin and charge currents,
occurring due to spin Coulomb drag in individual layers~\cite{weberscd,scd}
and due to spin Hall drag~\cite{shd} in electronic bilayers.

In zinc-blende semiconductor nanostructures the interplay between different
mechanisms of SOI can itself have crucial consequences. In the presence of
both Bychkov-Rashba (BR)~\cite{rashba} and Dresselhaus (D)~\cite{dresselhaus}
SOI the system possesses $C_{2v}$ symmetry. The BR coupling strength $\alpha 
$ depends largely on the asymmetry of structure while the D coupling $\beta $
vary mainly with the thickness of structure. In the special case when the BR
and D SOI strengths are adjusted~\cite{ganichev,giglberger} to be equal,
even higher $SU(2)$ symmetry occurs in the system~\cite{bernevig} and
various relaxation~\cite{schliemann} and optical~\cite{maytorena} properties
of the system turned out to be identical to those in the absence of SOI. A
remarkable demonstration of such suppression of SOI is the fresh
experimental realization~\cite{koralek} of the persistent spin-helix.

Another distinct manifestation of the interplay of BR and D mechanisms is
the SOI-induced anisotropy of single-particle spectrum, which modifies spin
relaxation and transport properties of the system~\cite{golub,stano,moser}.
Recently we have studied the influence of that anisotropy on the many-body
response of a 2DES~\cite{bavf2009}. Our calculations have revealed a fine
structure of the plasmon spectrum, which produces a striking asymmetric
doublet of the structure factor versus momentum orientation. The joint
action of BR and D SOI leads to dependence of the \textit{inter-chirality}
particle-hole continuum on direction. Thus, the plasmon propagation may be
free in one direction, but strongly damped in a different direction, where
the plasmon dispersion enters the particle-hole continuum. This creates a
possibility of directional plasmon filtering, potentially useful for both
spintronic and plasmonic devices.

In this Letter we study the many-body response of 2DES in
the presence of BR and D SOI in the static limit. By exploiting our recently
derived \textit{exact} formula for the Lindhard polarization function (PF),
we reveal a highly anisotropic behavior of the static dielectric function.
Particularly, the \textit{inter-chirality} transitions strongly modify the
singular behavior of the dielectric function. The sharp anisotropic Fermi
surface makes the position of singularities dependent on momentum
orientation and on the ratio of the BR and D SOI strengths, in addition to
the usual dependence on the magnitude of momentum $q$. We find that PF shows
a sharp cusp at $q<2k_{F}$ ($k_{F}$ is the Fermi wave vector) for the
momentum orientation along the $[1\bar{1}0]$ direction while in the
perpendicular $[110]$ direction, the singularity occurs at $q>2k_{F}$. Most
importantly, we observe that PF exhibits a \textit{doubly-singular }behavior$%
-$the singularities occur both at $q<2k_{F}$ and $q>2k_{F}$. As a direct
consequence of this, we find that\ the Friedel oscillations propagate with
two slightly different spatial frequencies and a novel \textit{beating
phenomenon of Friedel oscillations} takes place. At strictly equal SOI
strengths, $\alpha =\beta $, 
only the second singularity survives. The Friedel oscillations become
isotropic and the effect of SOI reduces to a simple renormalization of the
position of singularity. 

\begin{figure}[t]
\includegraphics[width=.55\linewidth]{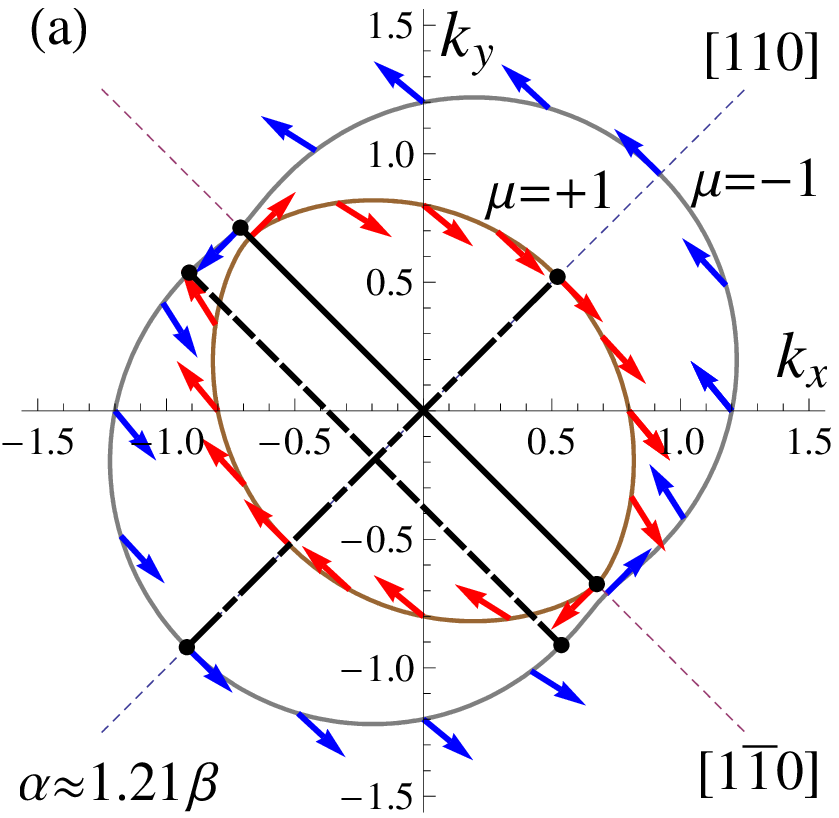} %
\includegraphics[width=.43\linewidth]{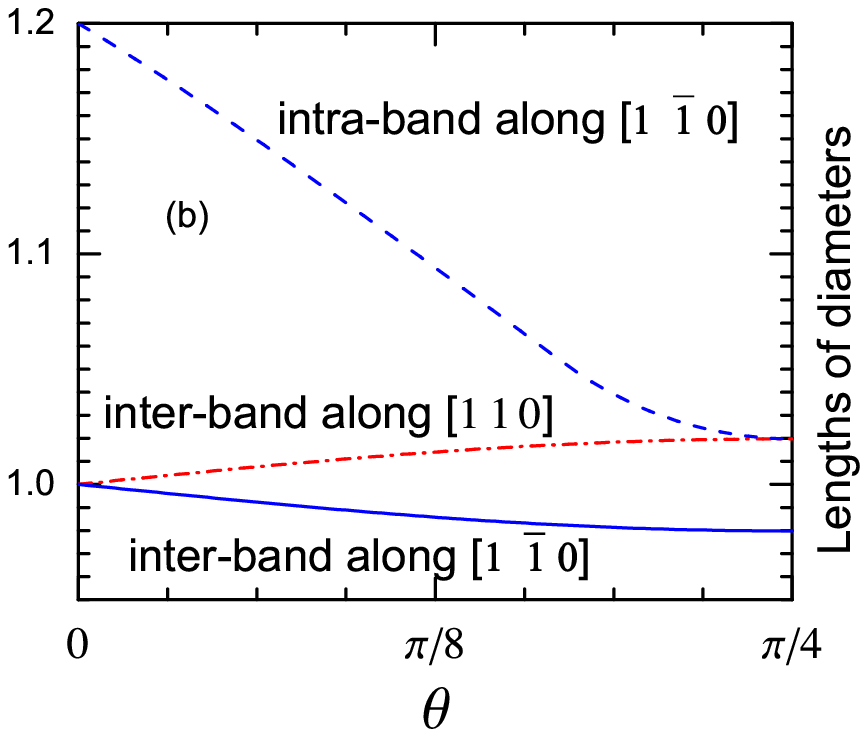}
\caption{(Color Online) (a) The Fermi contour in the presence of BR+D SOI.
Arrows indicate directions of spin. The thick solid and dashed lines are
diameters along the $[1\bar{1}0]$ direction, connecting maximally distant
points on the different chirality subbands and on the outer subband,
respectively. The thick dash-dotted line is the inter-subband diameter in $%
[110]$ direction. The BR and D strengths are related as $\protect\alpha =1.21%
\protect\beta $ with $\protect\rho =0.2k_{F}$. (b) The lengths of respective
diameters of the Fermi contour as a function of the parameter $\protect%
\theta $ for $\protect\rho =0.2k_{F}$.}
\label{FC}
\end{figure}

The Hamiltonian of BR and D SOI in quantum wells of zinc-blende structure,
grown on a $(001)$ surface, is $H_{\text{SOI}}=\alpha \left( \hat{\sigma}%
_{x}k_{y}-\hat{\sigma}_{y}k_{x}\right) +\beta \left( \hat{\sigma}_{x}k_{x}-%
\hat{\sigma}_{y}k_{y}\right) $ where $\hat{\sigma}_{x,y}$ are the Pauli
matrices, $\vec{k}$ is the in-plane electron momentum with its magnitude $k$
and polar angle $\phi _{\mathbf{k}}$. The eigenvectors of the Hamiltonian $%
H=H_{0}+H_{\text{SOI}}$ with $H_{0}=\vec{k}^{2}/2m^{\ast }$ ($m^{\ast }$ is
the electron effective mass and $\hbar =1$) are $\Psi _{\mu }(\vec{r})=\frac{%
1}{\sqrt{2\mathcal{A}}}\left( 
\begin{array}{c}
ie^{-i\varphi } \\ 
\mu%
\end{array}%
\right) e^{i\vec{k}\vec{r}}$. They correspond to the energy branches $E_{\mu
}(\vec{k})=\frac{1}{2m^{\ast }}\left[ \left( k+\mu \ \xi (\rho ,\theta ,\phi
_{\mathbf{k}})\right) ^{2}-\xi (\rho ,\theta ,\phi _{\mathbf{k}})^{2}\right]
,$ which are labeled by the \textit{chirality} $\mu =\pm 1$. Here $\mathcal{A%
}$ is the normalization area and the phase of spinor is given by $\varphi
(\alpha ,\beta ,\phi _{\mathbf{k}})=$Arg$[\alpha e^{i\phi _{\mathbf{k}%
}}+i\beta e^{-i\phi _{\mathbf{k}}}].$ The angle-dependent BR-D momentum is $%
\xi (\rho ,\theta ,\phi _{\mathbf{k}})=\rho \sqrt{1+\sin (2\theta )\sin
(2\phi _{\mathbf{k}})}$ where $\rho =m^{\ast }\sqrt{\alpha ^{2}+\beta ^{2}}$%
. The angle parameter $\theta $ is defined as $\tan \theta =\beta /\alpha $
and describes the relative strength of the BR and D SOI. The Fermi momenta
of the chirality subbands are also angle dependent: $k_{F}^{\mu }(\rho
,\theta ,\phi _{\mathbf{k}})=\sqrt{2mE_{F}+\xi (\rho ,\theta ,\phi _{\mathbf{%
k}})^{2}}-\mu \ \xi (\rho ,\theta ,\phi _{\mathbf{k}})$ where the total
carrier density $n$ determines the Fermi energy, $E_{F}=\left( \pi n-\rho
^{2}\right) /m^{\ast }$. Fig.~\ref{FC} shows the anisotropic Fermi contour
in the $\left( k_{x},k_{y}\right) $ plane.

In the static limit the dielectric function $\varepsilon (\vec{q})=1-v(q)\Pi
(\vec{q})$ where $v(q)=2\pi e^{2}/(\kappa _{0}q)F(qd)$ is the bare Coulomb
interaction with $\kappa _{0}$ the low frequency dielectric constant. The
form factor $F(qd)$ takes into account the transverse width $d$ of the
quantum well. It goes as $1-(1/3-5/4\pi ^{2})qd$ in the long wavelength
limit $qd\rightarrow 0$, and as $3/(4\pi ^{2}qd)$, in the opposite limit $%
qd\rightarrow \infty $. The exact PF $\Pi (\vec{q})$ can be expressed in
terms of the non-interacting Lindhard response function, $\Pi ^{0}(\vec{q})$%
, as $\Pi \left( \vec{q}\right) =\Pi ^{0}(\vec{q})\left[ 1-v(q)\left(
1-G_{+}\left( q\right) \right) \Pi ^{0}(\vec{q})\right] ^{-1}.$ Here the
\textquotedblleft charge-channel" local field factor $G_{+}\left( q\right) $~%
\cite{GV05} takes into account all electron correlations, related to the
vertex corrections beyond the random phase approximation. We neglect the
effect of SOI on $G_{+}$.

The static Lindhard PF in the presence of SOI is defined as a sum over the
indices $\mu $ and $\nu $, $\Pi ^{0}(\vec{q})=\sum_{\mu ,\nu =\pm 1}\Pi
_{\mu \nu }^{0}(\vec{q})$, where%
\begin{equation}
\Pi _{\mu \nu }^{0}(\vec{q})=\int \frac{d\vec{k}}{\left( 2\pi \right) ^{2}}%
\frac{f_{F}(E_{\mu }(\vec{k}))-f_{F}(E_{\mu \cdot \nu }(\vec{k}+\vec{q}))}{%
E_{\mu }(\vec{k})-E_{\mu \cdot \nu }(\vec{k}+\vec{q})}\mathcal{F}_{\nu
}\left( \vec{k},\vec{k}+\vec{q}\right) ~.  \label{eq10}
\end{equation}%
Here $\nu =\pm 1$ refers to the intra- and inter-subband contributions to $%
\Pi ^{0}(\vec{q})$ and $f_{F}(E_{\mu }(\vec{k}))$ are the Fermi distribution
functions. The form factor $\mathcal{F}_{\mathcal{\nu }}\left( \vec{k},\vec{k%
}+\vec{q}\right) $\ comes from the product of bare vertices and describes
the overlapping of spinor wave functions. It is given by $\mathcal{F}_{\nu
}\left( \vec{k},\vec{k}+\vec{q}\right) =\frac{1}{2}\left[ 1+\nu \cos \left(
\Delta \varphi _{\mathbf{q}}\right) \right] $ where we define $\Delta
\varphi _{\mathbf{q}}=\varphi (\alpha ,\beta ,\phi _{\mathbf{k}})-\varphi
(\alpha ,\beta ,\phi _{\mathbf{k}+\mathbf{q}})$. Notice that in contrast to
the case with pure BR or D SOI, here PF depends additionally on the momentum
orientation, $\phi _{\mathbf{q}}$. Recently in Ref.~\onlinecite{bavf2009} we
have derived an exact formula for the PF in the presence of BR+D SOI, which
is omitted here for the sake of brevity. The results, obtained from this
formula are in agreement with the previous classic result by Stern~\cite%
{stern} and findings by Pletyukhov and Gritsev~\cite{pletyukhov},
respectively, in the limits of vanishing SOI and of pure BR SOI.

\begin{figure}[t]
\includegraphics[height=.43\linewidth]{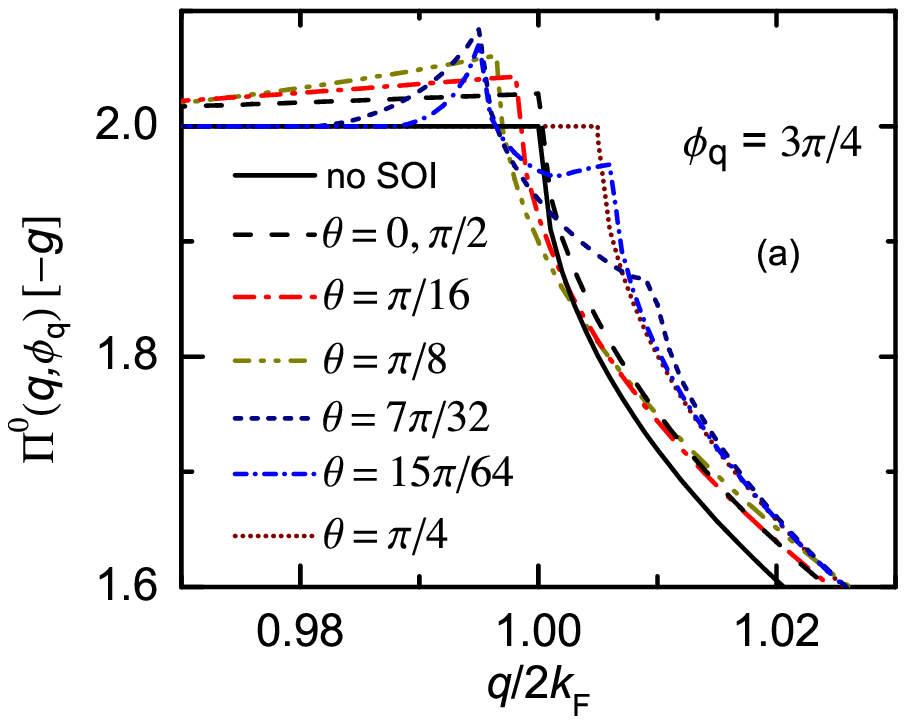} \includegraphics[height=.43%
\linewidth]{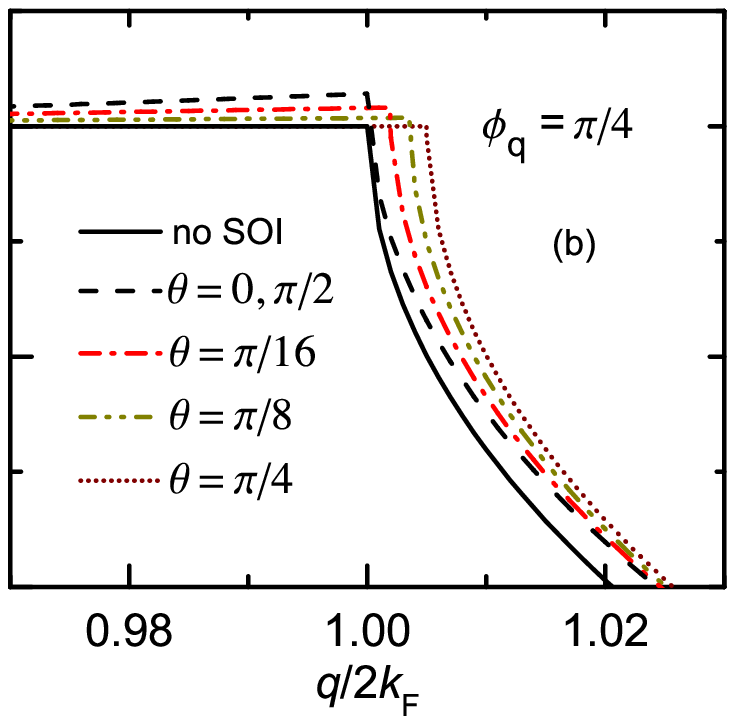}
\caption{(Color online) The static polarization function $\Pi ^{0}(\vec{q})$
as a function of the momentum $q$ for its two orientations: (a) $\protect%
\phi _{\mathbf{q}}=3\protect\pi /4$ and (b) $\protect\pi /4$. The different
curves correspond to different values of $\protect\theta $, shown on the
graph legends for $r=0.1$.}
\label{pf}
\end{figure}

Further by exploiting this general formula for PF~\cite{bavf2009}, we
calculate the static PF and analyze the modifications induced by the BR+D
SOI. We use the realistic materials parameters for InAs quantum wells by
taking $m^{\ast }=0.023m_{0}$, $\kappa _{0}=14.55$, and the transverse width
of the quantum well $d=15$ nm. We have also defined and will use in the 
following the dimensionless parameters $x=q/2k_{F}$ and $r=\rho /k_{F}$ with 
$k_{F}=\sqrt{2m^{\ast }E_{F}+\rho ^{2}}$. In Fig.~\ref{pf} we plot $\Pi ^{0}(\vec{q})$ 
in units of the density of states at the Fermi level $g=m^{\ast }/2\pi$ as a
function of $q$ for two different orientations of the momentum, $\phi _{%
\mathbf{q}}=\pi /4$ and $\phi _{\mathbf{q}}=3\pi /4$, and for several values
of the angle parameter $\theta $. The solid line represents PF in the
absence of SOI, $r=0$, while the dashed line corresponds to PF $\Pi
_{BRD}^{0}(\left\vert \vec{q}\right\vert )$ in the case of the pure BR or D
SOI, $\theta =0$ or $\pi /2$, respectively. All the curves of $\Pi ^{0}(\vec{%
q})$ show singular behavior at one or two values of $q$, determined by the
lengths of diameters in Fig.~\ref{FC}a and \ref{FC}b. Independent of $\theta 
$, the curves, which refer to $\phi _{\mathbf{q}}=3\pi /4$, exhibit a
singularity at the wave vectors $q_{1c}=2k_{F}\sqrt{1-r^{2}\sin 2\theta }%
<2k_{F}$ (cf. the solid line diameter in Fig.~\ref{FC}) with a maximum
polarizability at the singular point, $\Pi _{1}^{\max }$, exceeding the
maximum value of $\Pi _{BRD}^{0}(\left\vert \vec{q}\right\vert )$. At wave
vectors $q_{2c}>2k_{F}$ PF develops a \textit{second point} of
nonanalyticity with a local maximum $\Pi _{2}^{\max }<2$. Here $%
q_{2c}=\max_{\phi _{\mathbf{k}}}Q(\phi _{\mathbf{k}})$ with the function $%
Q(\phi _{\mathbf{k}})=\left\vert \left( \sin (\phi _{\mathbf{k}})-\cos (\phi
_{\mathbf{k}})\right) \left[ \overline{\xi }_{\mathbf{k}}+\sqrt{1+r^{2}\sin
\left( 2\phi _{\mathbf{k}}\right) \sin \left( 2\theta \right) }\right]
\right\vert $ (cf. the dashed line diameter in Fig.~\ref{FC}). For values of 
$\theta $ not far from $\pi /4$, the singularity at $q_{2c}$ becomes well
pronounced while the singularity at $q_{1c}$ turns into a sharp cusp. The
curves in Fig.~\ref{pf}, which refer to $\phi _{q}=\pi /4$, exhibit only one
singularity at the position $q_{3c}=2k_{F}\sqrt{1+r^{2}\sin 2\theta }>2k_{F}$
(cf. the dash-dotted line diameter in Fig.~\ref{FC}) and the polarizability
in the singular point varies within the $2<\Pi ^{0}(\vec{q})<\Pi
_{BRD}^{0}(\left\vert \vec{q}\right\vert )$\ window. In all singular points
PF is continuous and its derivative discontinuous. At the special values of $%
\theta =\pi /4$ and $3\pi /4$ when $\alpha =\pm \beta $, the form factor $%
\mathcal{F}_{\nu }$ ceases to depend on the wave vector $\vec{k}$ and
reduces to the Kronecker symbol. In this case we find that the effect of SOI
on the density response of a 2DES \textit{disappears} for any value of $\phi
_{\mathbf{q}}$. The only remaining modification reduces to a simple
renormalization of the isotropic Fermi wave vector, $k_{F}\rightarrow k_{Fc}=%
\sqrt{1+r^{2}}k_{F}$.

The electron density deviation, generated by the perturbation of a single
impurity, which is embedded at $R=0$ in the electron sheet in the $(x,y)$
plane, is determined by the static density response function via the relation%
\begin{equation}
\delta n(\vec{R})=\int \frac{d\vec{q}}{\left( 2\pi \right) ^{2}}e^{i\vec{q}%
\vec{R}}V_{i}(q)\Pi (\vec{q})  \label{eq18}
\end{equation}%
where $V_{i}(q)$\ is the Fourier transform of an isotropic impurity
potential. Since in the presence of BR+D SOI PF is anisotropic, we can
rewrite Eq.~\ref{eq18} as%
\begin{equation}
\frac{\delta n(\vec{R})}{n}=\frac{2}{\pi }\int_{0}^{\infty
}dxxV_{i}(x)\int_{0}^{2\pi }d\phi _{\mathbf{q}}e^{i\xi x\cos \left( \phi _{%
\mathbf{q}}-\phi _{\mathbf{R}}\right) }\Pi (x,\phi _{\mathbf{q}})
\label{eq19}
\end{equation}%
where $\xi =2k_{F}R=\left( 2\sqrt{2}/r_{s}\right) \left( R/a_{B}\right) $
and $\phi _{\mathbf{R}}$ is the polar angle of the vector $\vec{R}$, $r_{s}=%
\sqrt{2}/k_{F}a_{B}$\ the dimensionless constant of electron-electron
interaction$~$\cite{GV05}, $a_{B}$ the effective Bohr radius. At large
distances from the impurity, the integrand has a rapidly-varying phase, $\xi
x\cos \left( \phi _{\mathbf{q}}-\phi _{\mathbf{R}}\right) $ so the main
contribution to the integral comes from the point where the phase is
stationary and the exponential function $e^{i\xi x\cos \left( \phi _{\mathbf{%
q}-}\phi _{\mathbf{R}}\right) }$ oscillates less rapidly. The application of
the method of stationary phase to the integral over $\phi _{\mathbf{q}}$ in (%
\ref{eq19}) yields%
\begin{equation}
\frac{\delta n(\vec{R})}{n}\sim 4\sqrt{\frac{2}{\pi \xi }}\int_{0}^{\infty
}dx\sqrt{x}V_{imp}(x)\Pi (x,\phi _{\mathbf{R}})\cos \left( \frac{\pi }{4}%
-\xi x\right) ~.  \label{eq20}
\end{equation}%
Here we have used the relation $\Pi (x,0)=\Pi (x,\pi )$, which is the case
in the presence of BR+D SOI due to $C_{2v}$ symmetry. Thus, at large
distances from the impurity one can replace the second argument of PF $\phi
_{\mathbf{q}}$ by $\phi _{\mathbf{R}}$. 

In order to take the integration over $x$ we can exploit the
Riemann--Lebesgue lemma, which says that if a function oscillates rapidly
around zero then the integral of this function is small and the principal
contribution to the integral is determined by the integrand behavior in the
neighborhood of singular points. Therefore, one can reduce Eq.$~\ref{eq20}$
to the following asymptotic expression for the density deviation 
\begin{equation}
\frac{\delta n(\vec{R})}{n}\sim \frac{1}{g}\sum\limits_{i}\sqrt{\frac{2x_{ic}%
}{\pi \xi }}A(x_{ic})\int dx\delta \Pi ^{0}(x,x_{ic})\cos \left( \frac{\pi }{%
4}-\xi x\right)   \label{eq21}
\end{equation}%
where%
\begin{equation}
A(x_{ic})=\frac{4gV_{imp}(x_{ic})}{\left[ 1-v(x_{ic})\left( 1-G_{+}\left(
x_{ic}\right) \right) \Pi ^{0}(x_{ic},\phi _{\mathbf{R}})\right] ^{2}}
\label{eq22}
\end{equation}%
and $x_{ic}=x_{ic}\left( r,\theta ,\phi _{\mathbf{q}}\right) $ denotes the
position of the $i$th singularity\ of PF. The increment of PF $\delta \Pi
^{0}(x,x_{ic})=\Pi ^{0}(x,\phi _{\mathbf{R}})-\Pi ^{0}(x_{ic},\phi _{\mathbf{%
R}})$ near the singularity $x_{ic}$ can be represented as%
\begin{equation}
\delta \Pi ^{0}(x,x_{ic})\approx -g\vartheta \left[ \pm \left(
x-x_{ic}\right) \right] \frac{a_{_{i}}}{x}\left\vert
x^{2}-x_{ic}^{2}\right\vert ^{\alpha _{\pm ,i}}~  \label{eq23}
\end{equation}%
where $\vartheta \left( x\right) $ is the unit step function, the signs $\pm 
$\ corresponds to the function below ($x<x_{ic}$) and above ($x>x_{ic}$) the
singularity $x_{ic}$. The critical exponents $\alpha _{\pm ,i}$ and the
coefficients $a_{_{i}}$ describe the power law behavior and the maximum
polarizability at the singular points. In Eq.$~\ref{eq21}$ we have assumed
that the nonanalytic behavior of the interacting PF $\Pi (\vec{q})$ is
completely determined by its noninteracting part $\Pi ^{0}(\vec{q})$~\cite{GV05}. 
Substituting Eq.~\ref{eq23} into Eq.~\ref{eq21} and making use the Lighthill 
theorem~\cite{lighthill}, after the integration over $x$ we get%
\begin{equation}
\frac{\delta n(\vec{R})}{n}\sim -\frac{2A_{0}}{\sqrt{\pi \xi }}%
\sum\limits_{\pm ,i}a_{_{i}}\frac{\left( \alpha _{\pm ,i}\right) !}{\xi
^{1+\alpha _{\pm ,i}}}\cos \left[ \xi x_{ic}+\frac{\pi }{2}\left( \alpha
_{\pm ,i}+\frac{1}{2}\right) \right] ~.  \label{eq24}
\end{equation}

As we have already discussed, PF can qualitatively change its singular
behavior, depending on the ratio of the BR and D SOI coupling strengths as
well as on the momentum orientation. Accordingly, the Friedel oscillations,
given by Eq.$~$\ref{eq24}, can exhibit completely new features. For
instance, at $\theta =\pi /8$ PF for two orthogonal orientations, $\phi _{%
\mathbf{q}}=\pi /4$ and $\phi _{\mathbf{q}}=3\pi /4$, shows singularities,
respectively, at the positions $x_{+c}$ and $x_{-c}$ with $x_{\pm c}\approx
1\pm \delta _{c}.$ It is clear that the phase difference between these two
orientations is about $\left( x_{+c}-x_{-c}\right) \xi =\left( 4\sqrt{2}%
/r_{s}\right) \delta _{c}\left( R/a_{B}\right) $ and can result in a
striking difference in the behavior of Friedel oscillations at the distance $%
R/a_{B}$ of the order of $r_{s}/4\sqrt{2}\delta _{c}$. In InAs samples with
the electron density $n=10^{16}$ m$^{-2}$, we have $r_{s}\approx 0.12$ and
for $r=0.1$ taking $\delta _{c}=0.5\times 10^{-2}$ we obtain that the
Friedel oscillations in the $\phi _{\mathbf{R}}=\pi /4$ and $\phi _{\mathbf{R%
}}=3\pi /4$ directions are in \textit{antiphase} at the distance of the
order of $R\lesssim 5a_{B}$.

\begin{figure}[t]
\includegraphics[width=.49\linewidth]{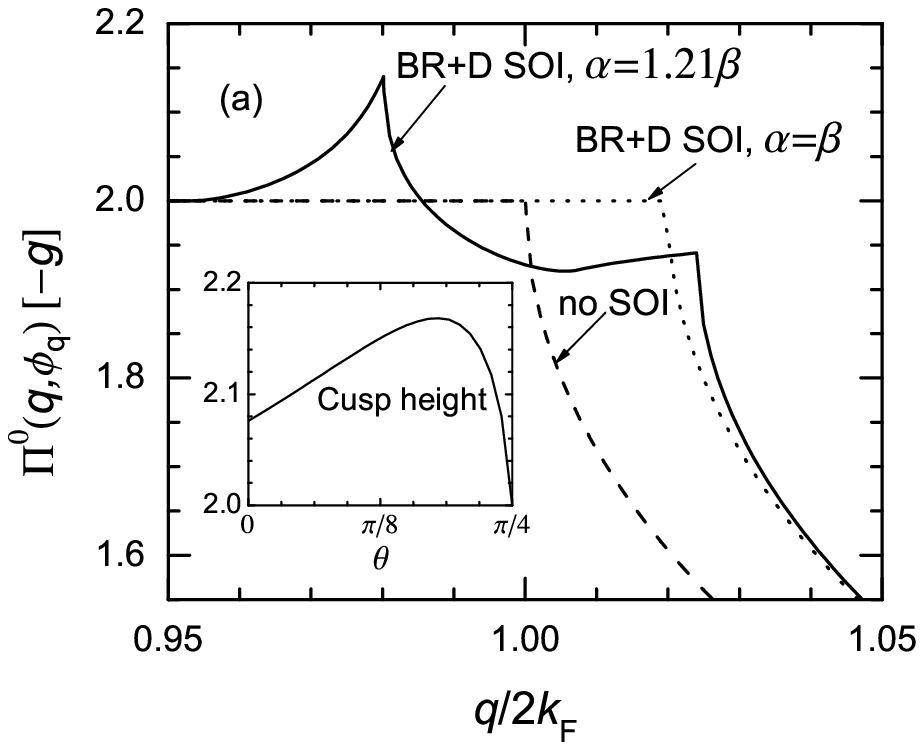} \hfil
\includegraphics[width=.49\linewidth]{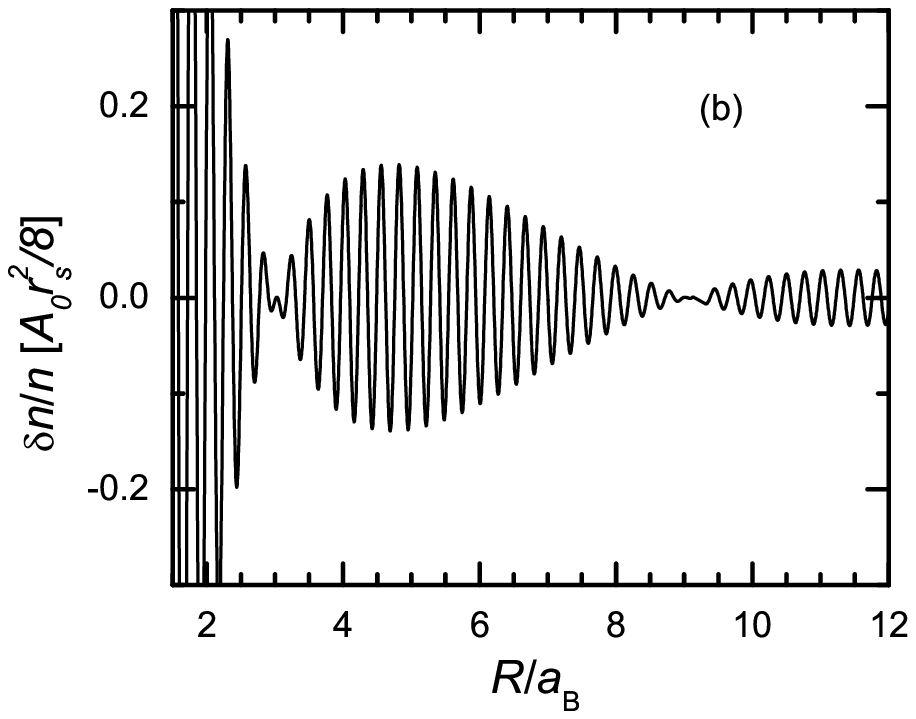}
\caption{(a) The doubly-singular polarization function for $\protect\alpha=
1.21\protect\beta$ with $\protect\rho =0.2k_{F}$ (solid line). Inset shows
the height of cusp vs $\protect\theta$. The dotted and dashed curves
represent $\Pi ^{0}(\vec{q})$ for $\protect\alpha=\protect\beta$ with $%
\protect\rho =0.2k_{F}$ and $\protect\rho=0$, respectively. (b) Beating of
Friedel oscillations, induced by the doubly-singular behavior of $\Pi ^{0}(%
\vec{q})$ in (a). The electron density $n=2\times 10^{16}$ m$^{-2}$.}
\label{bfo}
\end{figure}

Another interesting effect appears when PF exhibits the doubly-singular
behavior (cf. the curves with $\theta =7\pi /32$ or $7\pi /32$ in Fig.~\ref%
{pf}). In Fig.~\ref{bfo}a we illustrate the doubly-singular behavior of PF
separately for $r=0.2$. Inset shows the height of the cusp at $q=q_{1c}$
(cf. the solid curve in Fig.~\ref{FC}b) versus $\theta $. As seen, $\Pi
_{1}^{\max }=\Pi _{BRD}^{0}$ at $\theta =0$ and increases with $\theta $ up
to its maximum $\Pi _{1}^{\max }\approx 2.17$ at about $\theta =0.18\pi $.
With a further increase of $\theta $, $\Pi _{1}^{\max }$ drops sharply to
its value in the absence of SOI, $\Pi _{1}^{\max }=2$ at $\theta =\pi /4$.
By fitting numerically the singularities of the \textit{exact} PF, we have
established with numerical certainty that the critical exponents in the
power law behavior in Eq.~\ref{eq23} are (cf. Ref.~\onlinecite{pletyukhov}): 
$\alpha _{-,1,2}=1$ and $\alpha _{+,1,2}=1/2$, respectively, for the left-
and right-side of both singularities at $x_{1c}=1-\delta _{1c}$ and $%
x_{2c}=1+\delta _{2c}$. Hence, in the limit of large $\xi $ the
contributions to $\delta n(\vec{R})$, coming from the left-side neighborhood
of both\ singularities, are small. Taking into account also that the smooth
functions $A(x_{ic})$ and $a_{_{i}}$ do not depend on the large parameter $%
\xi $, one can approximate $a_{_{i}}A(x_{ic})\approx 2A(1)\equiv A_{0}$.
Thus, the density deviation at large distances from the impurity can be
reduced to the following simple form%
\begin{equation}
\frac{\delta n(\vec{R})}{n}\sim \frac{A_{0}}{\xi ^{2}}\left[ \sin \left(
1-\delta _{1c}\right) \xi +\sin \left( 1+\delta _{1c}\right) \xi \right] ~.
\end{equation}%
It is clear that due to the existence of two singularities at $%
x_{1c}=1-\delta _{1c}$ and $x_{2c}=1+\delta _{2c}$, the Friedel oscillations 
$\sin \left( 1-\delta _{1c}\right) \xi $ and $\sin \left( 1+\delta
_{1c}\right) \xi $ propagate with two quite close spatial frequencies and a 
\textit{beating} phenomenon of Friedel oscillations can be observed at the
beat frequency $\sqrt{2}\left( \delta _{1c}+\delta _{2c}\right) /r_{s}$.
Fig.~\ref{bfo}b illustrates the first two destructive interferences of the
Friedel oscillations that occur at $R\sim 3a_{B}$ and $9a_{B}$. Taking into 
account higher order terms in $\xi $ and $\rho $ will partially smooth the
interference picture, however, the beating of Friedel oscillations as a
distinct modulation of the density deviation is a stable feature of systems
with BR+D SOI and should be observable in experiment. Notice also that in 
samples with a stronger SOI such as HgTe quantum wells, the separation 
$\delta _{1c}+\delta _{2c}$ between the singularities increases, which will 
essentially facilitate the experimental detection of the destructive 
interferences.

In conclusion, we have calculated the static response of a 2DES in the
presence of joint BR and D SOI. We find that one of the main modifications
is the induced shift of the singularity position of the static PF, which is
in opposite directions for orthogonal momentum orientations. This results in
a strong anisotropy of the Friedel oscillations. More interestingly, we have
shown that in certain situations PF exhibits a \textit{doubly-singular}
behavior, which generates a novel phenomenon -- \textit{the beating of
Friedel oscillations}. This intriguing prediction exemplifies how usually
weak SOI can generate a qualitatively new and physically robust occurrence
as a measurable signature of the many-body response of a 2DES.\newline

We acknowledges support from EU Grant PIIF-GA-2009-235394 (SMB), SFB Grant 
No. 689, and NSF Grant No. DMR-0705460 (GV).

\end{document}